\documentclass[12pt]{elsart3p}
\usepackage[dvips]{graphicx,psfrag}

\usepackage{float}
\usepackage{booktabs}
\usepackage{amsmath, amssymb}
\usepackage{bm}
\usepackage{ulem} 
\usepackage[usenames]{color}

\def\be{\begin{equation}}
\def\ee{\end{equation}}
\def\beal{\begin{align}}
\def\enal{\end{align}}
\def\bc{\begin{center}}
\def\ec{\end{center}}
\def\trm{\textrm}

\def\GE{G_{\trm{E}}}
\def\GM{G_{\trm{M}}}

\def\EMSR{\langle r^{2} \rangle _{\trm{E}}}
\def\MMSR{\langle r^{2} \rangle _{\trm{M}}}

\def\KbarN{\bar{K} N}
\def\Kmp{K^{-} p}
\def\Kzn{\bar{K}^{0} n}
\def\pzL{\pi ^{0} \Lambda}
\def\pzS{\pi ^{0} \Sigma ^{0}}
\def\etL{\eta \Lambda}
\def\etS{\eta \Sigma ^{0}}
\def\ppS{\pi ^{+} \Sigma ^{-}}
\def\pmS{\pi ^{-} \Sigma ^{+}}
\def\KpX{K^{+} \Xi ^{-}}
\def\KzX{K^{0} \Xi ^{0}}

\begin{document}
\begin{frontmatter}

\title{Electromagnetic mean squared radii of 
$\Lambda (1405)$ in chiral dynamics}

\author[KYOTO]{T. Sekihara},
\author[YITP,TUM]{T. Hyodo},
\author[YITP]{D. Jido}

\address[KYOTO]{Department of Physics, Kyoto University, Kyoto 606-8502,
 Japan}
\address[YITP]{Yukawa Institute for Theoretical Physics, Kyoto
 University, Kyoto 606-8502, Japan}
\address[TUM]{Physik-Department, Technische Universit\"at M\"unchen,
 D-85747 Garching, Germany}

\begin{abstract}
The electromagnetic mean squared radii, $\EMSR$ and 
$\MMSR$, of $\Lambda (1405)$ 
are calculated in the chiral unitary model. We describe the 
excited baryons as dynamically generated resonances 
in the octet meson and octet baryon scattering. 
We evaluate values of $\EMSR$ and $\MMSR$ for 
the $\Lambda (1405)$ on the resonance pole and obtain their complex values. 
We also consider $\Lambda (1405)$ obtained by neglecting 
decay channels. For the latter case, we obtain 
negative and larger absolute electric mean squared radius 
than that of typical ground state baryons. This implies 
that $\Lambda (1405)$ has structure that $K^{-}$ is widely spread 
around $p$. 

\end{abstract}


\end{frontmatter}

\section{Introduction}

The structure of the $\Lambda(1405)$ has been a long-standing 
issue. The $\Lambda (1405)$ has 
been considered as a quasi-bound state of anti-kaon and 
nucleon ($\KbarN$) 
system~\cite{Dalitz:1960du,Dalitz:1967fp,Wyld:1967}, before QCD 
was established.
Recent theoretical investigations have also suggested that the 
$\Lambda (1405)$ is well described as a dynamically generated resonant 
state of meson-baryon scattering with $I=0$ and $S=-1$
based on chiral dynamics in 
coupled channel 
approach~\cite{Kaiser:1995eg,Oset:1997it,Oller:2000fj,Oset:2001cn,Lutz:2001yb,Jido:2003cb,Hyodo:2008xr}, 
which is so-called chiral unitary model (ChUM). 
This model has successfully reproduced cross sections of $\Kmp$ to
various channels and also the mass spectrum of the $\Lambda (1405)$ 
resonance below the $\KbarN$ 
threshold~\cite{Kaiser:1995eg,Oset:1997it,Oller:2000fj,Lutz:2001yb,Jido:2002zk}, 
giving two states for 
the $\Lambda (1405)$ found as poles of the scattering amplitude 
in the complex energy 
plane~\cite{Oller:2000fj,Jido:2003cb,Fink:1989uk}. 
Photo-properties of $\Lambda (1405)$ have been investigated 
in the chiral unitary approach in 
Refs.~\cite{Kaiser:1996js,Nacher:1999ni,Jido:2002yz,Geng:2007hz}.

Most of excited baryons can be described by simple constituent 
quark models~\cite{Isgur:1978xj}. The $\Lambda (1405)$ is well known 
as one of the exceptions, so the success of the hadronic molecule 
picture is in this sense reasonable. 
Recent study of the $\Lambda (1405)$ based on the $N_c$ scaling in 
ChUM~\cite{Hyodo:2007np} indicates the dominance of the non-$qqq$ 
component in the $\Lambda (1405)$, and Ref.~\cite{Hyodo:2008xr} 
also suggests that 
the  $\Lambda (1405)$ is described predominantly by meson-baryon dynamics. 
The understanding of the structure of the $\Lambda (1405)$ is also relevant 
for the $\KbarN$ phenomenology, since binding energy of $\KbarN$ system 
in the $\Lambda (1405)$ plays an important role for the study of the kaonic 
nuclei~\cite{Akaishi:2002bg,Yamazaki:2007cs,Hyodo:2007jq,Dote:2008in}. 
Toward experimental verification, it is desirable to study some quantity
which characterizes the molecule structure of $\Lambda (1405)$
\cite{Jido:2002yz}. 

One expects that the $\Lambda (1405)$ as meson-baryon 
quasi-bound molecule with small binding energy has 
a larger size than typical ground state baryons dominated by a 
genuine quark state. If this is the case, the form factor of the 
$\Lambda (1405)$ falls off more rapidly than that of the nucleon and the 
production cross section of the $\Lambda (1405)$ has large energy 
dependence. In this work, we estimate the electromagnetic 
mean squared radii of the $\Lambda (1405)$ based on 
the ChUM. Such electromagnetic properties of the $\Lambda (1405)$ 
could be obtained in photon-induced production 
experiments~\cite{Exp:2007}.

\section{$\Lambda (1405)$ in chiral unitary model}
\label{sect:foundations}

In ChUM, the $\Lambda(1405)$ is described in $s$-wave meson-baryon scattering 
amplitudes with the strangeness $S=-1$ and charge $Q=0$ obtained by solving the 
Bethe-Salpeter (BS) equation, 
\be
T_{ij}(\sqrt{s}) = V_{ij}
+ \sum _{k} V_{ik} G_{k} T_{kj} ,
\label{eq:BSEq}
\ee
in which the on-shell factorization leads to an 
algebraic solution~\cite{Oller:2000fj}.
In Eq.~(\ref{eq:BSEq}), $V_{ij}$ is $s$-wave meson-baryon
interaction kernel and $G_{k}$ is loop integral 
of the meson-baryon system. They are  functions 
of the center-of-mass energy, $\sqrt{s}$, in matrix form 
of the meson-baryon channel ($i,j=\Kmp$, $\Kzn$, $\pzL$, $\pzS$, 
$\etL$ , $\etS$, $\ppS$, $\pmS$, $\KpX$ and $\KzX$). 
The interaction kernel $V_{ij}$ is given by the leading 
order chiral Lagrangian~\cite{Gasser:1984gg,Pich:1995bw}, which 
is known as Weinberg-Tomozawa term 
\be
V_{ij} = 
- \frac{C_{ij}}{4 f^{2}}(2 \sqrt{s} - M_{i} - M_{j}), 
\label{eq:Vij}
\ee
with an averaged meson decay constant $f = 1.123 f_{\pi}$, 
$f_{\pi}=93.0 \, \trm{MeV}$. 
This expression is obtained by applying the non-relativistic 
reduction for baryons. 
The coefficient $C_{ij}$ is fixed  by the SU(3) group structure of the  
interaction and its explicit value is given in~\cite{Oset:1997it}. 
The loop integral $G_{k}(\sqrt{s})$ is 
evaluated using the dimensional regularization:
\begin{align}
& G_{k} (\sqrt{s}) = 2 i M_{k} \int 
\frac{d ^{4} q_{1}}{(2 \pi)^{4}} 
\frac{1}{q_{1}^{2} - m_{k}^{2}} 
\frac{1}{(P - q_{1})^{2} - M_{k}^{2}}
 \nonumber \\
& = \frac{2 M_{k}}{16 \pi ^{2}} \Bigg[ a_{k} (\mu)
 + \ln \frac{M_{k}^{2}}{\mu ^{2}} 
 + \frac{m_{k}^{2} - M_{k}^{2} + s}{2 s} 
\ln \frac{m_{k}^{2}}{M_{k}^{2}}
 \nonumber \\
& \quad \phantom{+} + \frac{q_{k}}{\sqrt{s}} 
\Bigl( \ln (s - M_{k}^{2} + m_{k}^{2} + 2 q_{k} \sqrt{s}) 
\nonumber \\
 & \quad \phantom{+ (\frac{q_{k}}{\sqrt{s}})} 
+ \ln (s + M_{k}^{2} - m_{k}^{2} + 2 q_{k} \sqrt{s}) \nonumber \\
 & \quad \phantom{+ (\frac{q_{k}}{\sqrt{s}})} 
- \ln (- s + M_{k}^{2} - m_{k}^{2} + 2 q_{k} \sqrt{s}) \nonumber \\
 & \quad \phantom{+ (\frac{q_{k}}{\sqrt{s}})} 
- \ln (- s - M_{k}^{2} + m_{k}^{2} + 2 q_{k} \sqrt{s})\Bigr)
\Bigg] , 
\label{eq:loop}
\end{align}
where the center-of-mass 
momentum of the two-body system is given by
\be
q_{k} \equiv 
\sqrt{\frac{(s-M_{k}^{2}+m_{k}^{2})^{2} - 4 s m_{k}^{2}}{4 s}} .
\label{eq:q_k}
\ee
The subtraction constants $a_{k}(\mu)$ in Eq.~(\ref{eq:loop}) with the 
regularization scale $\mu$ are free parameters in this model. 
These constants are phenomenologically fixed so as to 
reproduce the threshold behavior of the scattering amplitudes~\cite{Oset:2001cn}:  
\be
\begin{split}
& a_{\KbarN} = -1.84, \phantom{{}_{\etL}} \! \! 
a_{\pi \Sigma} = -2.00, \phantom{{}_{\eta \Sigma}} \! \! 
a_{\pi \Lambda} = -1.83, \\ &
a_{\eta \Lambda} = -2.25, \phantom{{}_{\KbarN}} \! \! 
a_{\eta \Sigma} = -2.38, \phantom{{}_{\pi \Lambda}} \! \! 
a_{K \Xi} = -2.67, 
\end{split}
\label{eq:subtconst}
\ee
with $\mu=630 \, \trm{MeV}$.
In the present model, the $\Lambda(1405)$ is dynamically generated 
in the obtained BS scattering amplitude without introducing any explicit 
pole terms in the interaction kernel. 

\begin{figure}[t]
\bc
 \begin{tabular*}{200pt}{@{\extracolsep{\fill}}cc}
    \includegraphics[scale=0.145]{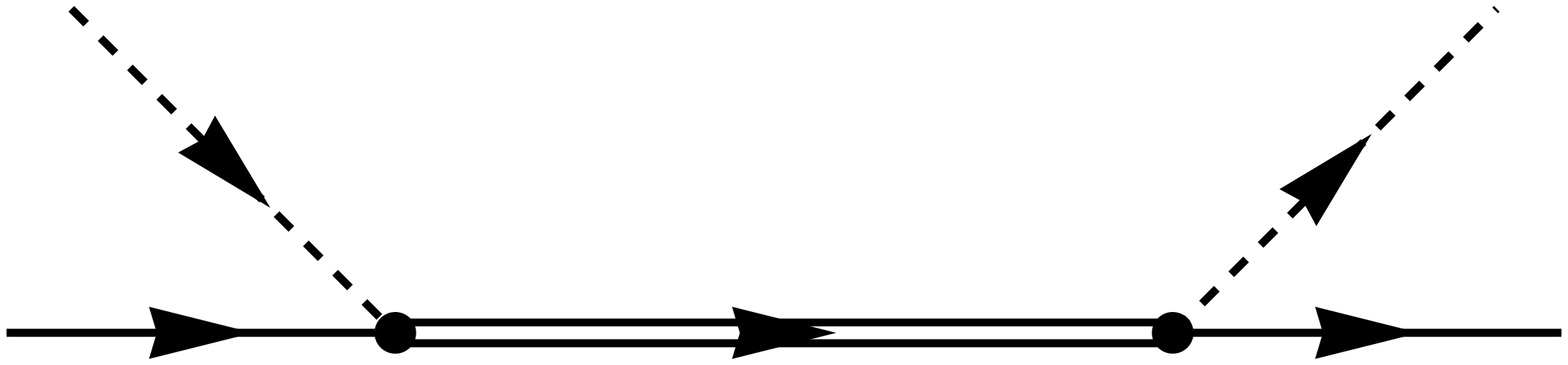} &
    \includegraphics[scale=0.145]{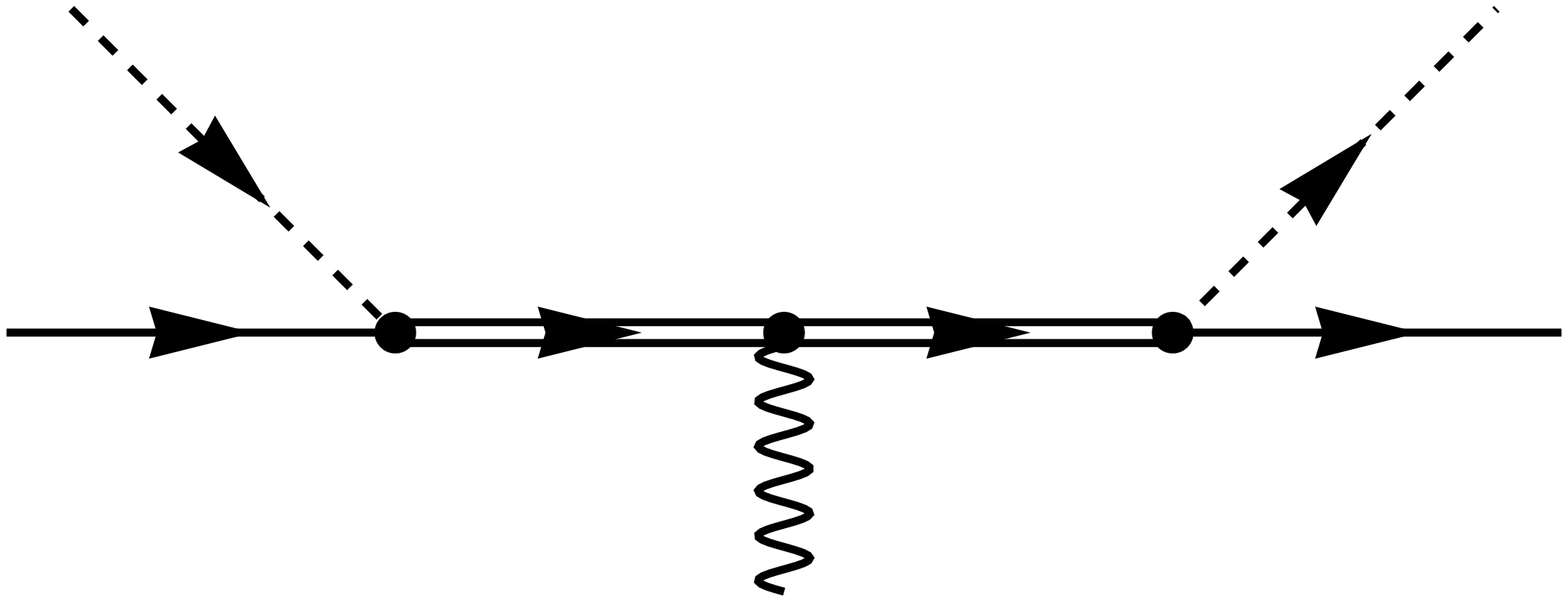}\\
    (a) $T_{ij}$ & (b) $T_{\gamma ij}^{\mu}$ 
 \end{tabular*}
\ec
\caption{Scattering amplitudes close to the pole of 
the excited baryon. The double lines correspond to the excited baryon.}
\label{fig:resonance}
\end{figure}

The excited baryon is expressed by a pole of the scattering 
amplitudes in complex energy plane. The $s$-wave scattering amplitude 
can be approximated as, 
\begin{equation}
- i T_{ij} 
  \approx (- i g_{i}) \frac{i}{\sqrt{s} - z_{\trm{H}}} (- i g_{j}), 
\label{eq:T_mat} 
\end{equation}
close to the resonance energy, as shown in Fig.~\ref{fig:resonance}(a). 
The real and imaginary parts of the pole position $z_{\trm{H}}$ 
express the mass and 
the half-width of the excited baryon, respectively, and the residues
of the pole, $g_{i}$ and $g_{j}$, represent coupling strengths of the 
excited baryon to the meson-baryon channels. In the present model, 
two poles in the BS amplitude are found in energies of 
the $\Lambda(1405)$ as 
($z_{1} = 1390-66i \, \trm{MeV}$) and 
($z_{2} = 1426-17i \, \trm{MeV}$)~\cite{Jido:2003cb}. 
It has been reported in Ref.~\cite{Borasoy:2005ie} that the position 
of the lower pole $z_{1}$ is dependent on details of model parameters, 
whereas that of the higher pole $z_{2}$ shows little dependence.

\section{Form factors of excited baryons}
\label{sect:photon}

\begin{figure*}[!ht]
 \bc
 \begin{tabular*}{\textwidth}{@{\extracolsep{\fill}}ccc}
    \includegraphics[scale=0.145]{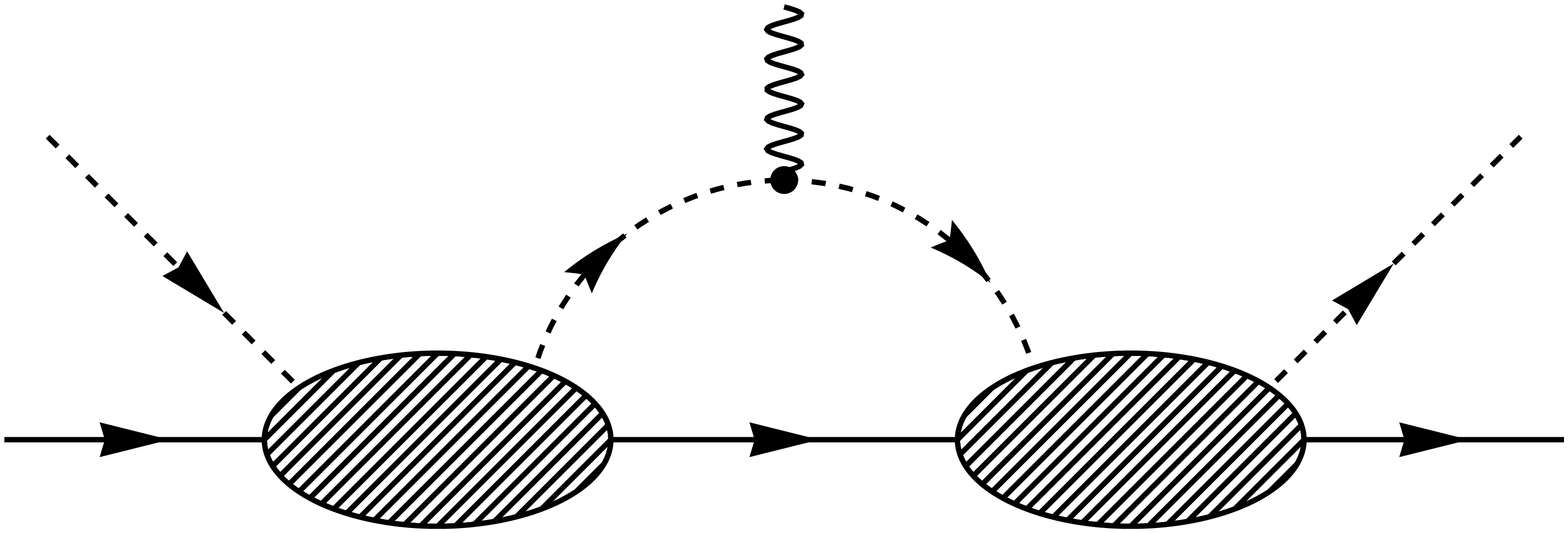} &
    \includegraphics[scale=0.145]{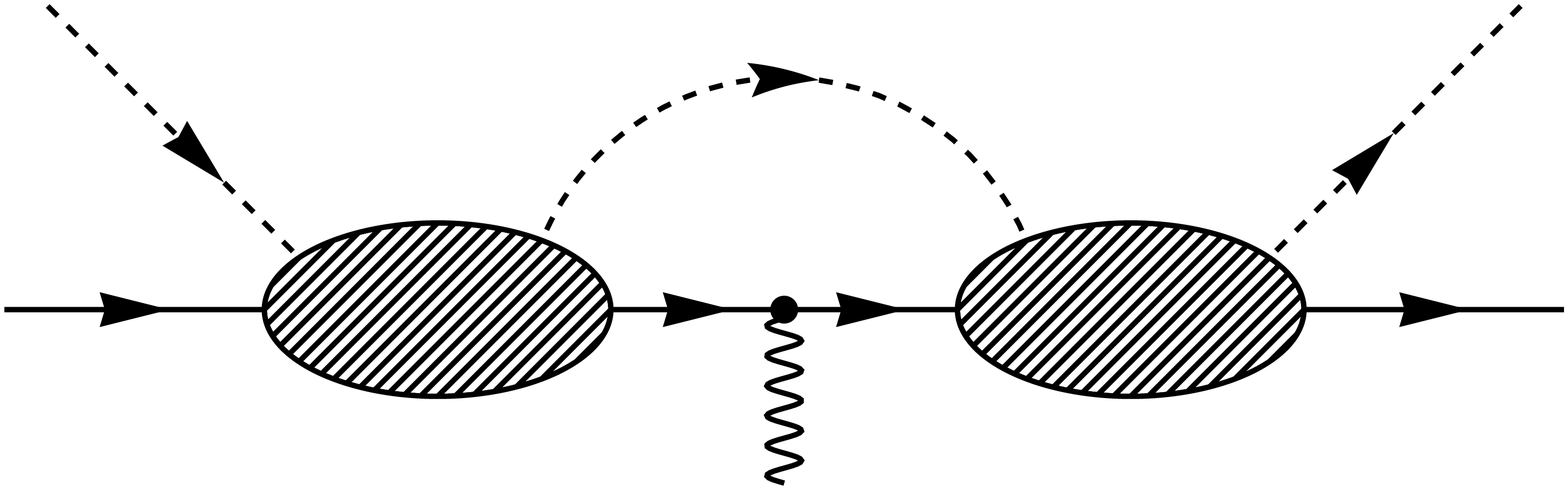} &
    \includegraphics[scale=0.145]{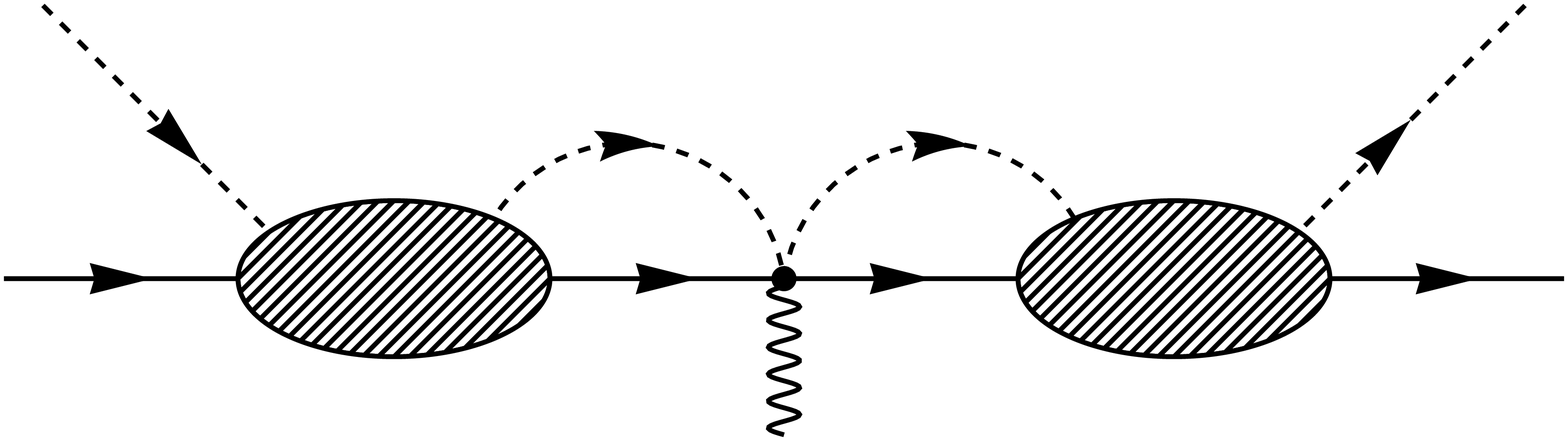} \\
    $T_{\gamma 1}^{\mu}$ &
    $T_{\gamma 2}^{\mu}$ &
    $T_{\gamma 3}^{\mu}$
 \end{tabular*}
 \ec
 \caption{Diagrams for the form factors of the excited baryon. 
The shaded ellipses represent the BS amplitude.}
 \label{fig:Tgamma}
\end{figure*}%

In this section, we discuss the formulation to evaluate the 
electromagnetic form factors and the mean squared radii of 
excited baryons described by the BS amplitudes. 
First of all, let us define the electromagnetic form factors of 
an excited baryon with spin 1/2, $H^{\ast}$, as matrix elements
of the electromagnetic current $J_{\trm{EM}}^{\mu}$ in the 
Breit frame~\cite{Jido:2002yz}:
\be
\big \langle H^{\ast} \big | J_{\trm{EM}}^{\mu} \big | H^{\ast}
\big \rangle _{\trm{Breit}}
\equiv \left( \GE (Q^{2}), \, 
\GM (Q^{2}) 
\frac{i \bm{\sigma} \times \bm{q}}{2 M_{\trm{p}}} 
\right) ,
\label{eq:definition}
\ee
with the electric and magnetic form factors, $\GE (Q^{2})$ and 
$\GM (Q^{2})$, the virtual photon momentum $q^{\mu}$, $Q^{2}=-q^{2}$ 
and the Pauli matrices $\sigma ^{a} \, (a=1, \, 2, \, 3) $. 
The magnetic form factor $\GM (Q^{2})$ is normalized as the nuclear 
magneton $\mu _{\trm{N}}=e/(2 M_{\trm{p}})$ with 
the proton charge $e$ and mass $M_{\trm{p}}$. 
From these form factors, electromagnetic mean squared radii, 
$\EMSR$ and $\MMSR$, are calculated by
\begin{align}
& \EMSR \equiv - 6 \left . 
\frac{d \GE}{d Q^{2}} \right | _{Q^{2}=0}, \\ 
& \MMSR \equiv 
- \frac{6}{G_{\trm{M}}(0)} \frac{d G_{\trm{M}}}
{d Q^{2}} \bigg |_{Q^{2}=0} .
\end{align}

To extract the matrix elements of the electromagnetic current 
from the BS amplitudes, we consider 
the scattering amplitude for the $MB \gamma ^{\ast} \rightarrow 
M^{\prime}B^{\prime}$ process, $T^{\mu}_{\gamma ij}$, 
which is microscopically calculated by attaching the photon to 
every place of the constituent meson-baryon components 
in the BS amplitude~\cite{Nacher:1999ni,Jido:2002yz,Borasoy:2005zg}. 
The matrix elements of the excited baryon can be 
expressed as residues of the double pole of the 
$MB \gamma ^{\ast} \rightarrow M^{\prime}B^{\prime}$ amplitude. 
Close to the pole of the excited baryon, 
as shown in Fig.~\ref{fig:resonance}(b),
the amplitude $T^{\mu}_{\gamma ij}$ is parametrized by  
\begin{align}
& - i T_{\gamma ij}^{\mu} ( \sqrt{\mathstrut s^{\prime}}, \, \sqrt{\mathstrut s} ) 
\nonumber \\
& \approx (- i g_{i}) \frac{i}{\sqrt{\mathstrut s^{\prime}} - z_{\trm{H}}} 
\big \langle H^{\ast} \big | i J_{\trm{EM}}^{\mu} \big | H^{\ast}
\big \rangle \frac{i}{\sqrt{\mathstrut s} - z_{\trm{H}}} 
(- i g_{j}) ,
\label{eq:T_gamma_mat}
\end{align}
where $s^{\prime}=P^{\prime \mu}P_{\mu}^{\prime}$ and $s=P^{\mu}P_{\mu}$, with 
incoming and outgoing momenta of excited baryon, $P^{\mu}$ and 
$P^{\prime \mu}$, respectively. 
Combining Eqs.~(\ref{eq:T_mat}) and (\ref{eq:T_gamma_mat}), 
the matrix elements can be evaluated as 
residue of the double pole at 
$\sqrt{\mathstrut s}=\sqrt{\mathstrut s^{\prime}}=z_{\trm{H}}$ 
as discussed in \cite{Jido:2002yz}: 
\be
\big \langle H^{\ast} \big | J_{\trm{EM}}^{\mu} \big | H^{\ast}
\big \rangle = 
\trm{Res} \left[ - 
\frac{T_{\gamma ij}^{\mu}(\sqrt{\mathstrut s^{\prime}}, \, \sqrt{\mathstrut s})}
{T_{ij} (\sqrt{\mathstrut s})} \Bigg |_{\sqrt{s} \rightarrow z_{\trm{H}}} \right] .
\label{eq:ResTT}
\ee
Here we calculate residue of $\sqrt{s^{\prime}}=z_{\trm{H}}$ by ``$\trm{Res}$'' 
in the right-hand side. 
Note that this evaluation is free from the non-resonant background, 
since the values are calculated just on the pole of the excited 
baryon. 

Thus, we have shown that the electromagnetic mean squared
radii of the excited baryon can be obtained, once the
scattering amplitude $MB \gamma ^{\ast} \rightarrow M^{\prime}B^{\prime}$
is calculated.

\section{Evaluation of the form factors}
\label{sect:EvalFF}

We calculate the scattering amplitude of the $MB\gamma^{\ast} \rightarrow M^{\prime} B^{\prime}$
in the chiral unitary approach, in which the amplitude for the 
$MB \rightarrow M^{\prime}B^{\prime}$ is given by multiple scattering of the meson and baryon. 
Thus, the photon couples to the $\Lambda(1405)$ through the constituent mesons
and baryons. 
The calculation should be performed in a gauge-invariant way,
which ensures the correct normalization of the electric form 
factor of the excited baryon, $\GE(Q^{2}=0)=Q_{\trm{H}}$.
Following the method of the gauge-invariant calculation for 
unitarized amplitudes proposed in Ref.~\cite{Borasoy:2005zg}, 
{we take three relevant diagrams shown in Fig.~\ref{fig:Tgamma},
which have the double pole for the excited baryon.  The other
diagrams do not contribute to the electromagnetic form factors 
at the resonance energy $\sqrt s = z_{\trm{H}}$.}
{Since the pole position is gauge invariant, these three contributions 
are enough for the gauge-invariant
form factors:}
\be
T_{\gamma ij}^{\mu} = T_{\gamma 1 ij}^{\mu} + 
T_{\gamma 2 ij}^{\mu} + T_{\gamma 3 ij}^{\mu} .
\label{eq:Amp_gauge}
\ee
Using this amplitude $T_{\gamma ij}^{\mu}$ with photon couplings to 
mesons and baryons which we will discuss later, 
we  obtain the gauge-invariant form factors through Eq.~(\ref{eq:ResTT}).
A proof of the gauge invariance of the form factors is given 
in Appendix A by using Ward-Takahashi identity. 

The elementary couplings of the photon to the meson and baryon
are given by imposing gauge invariance to the chiral effective 
theory in a consistent way
with the description of $\Lambda(1405)$.
In the present model, the $\Lambda(1405)$ is described by 
a infinite sum of the loop
function $G$ and the $s$-wave Weinberg-Tomozawa interaction $V$, 
with non-relativistic formulation for the baryons. 
We use the minimal coupling scheme for the photon couplings to
the meson and baryon appearing in the BS amplitude. This procedure 
automatically implements gauge invariance in a consistent way to 
the original  BS amplitude. We also have the anomalous magnetic couplings 
for the $BB\gamma$ and $\gamma BB^{\prime} MM^{\prime}$ couplings.
These couplings are given by the chiral perturbation theory as done
in Ref.~\cite{Jido:2002yz}. Finally the elementary electric and magnetic couplings, 
$ V_{\trm{M}_{i}}^{\mu}$, $ V_{\trm{B}_{i}}^{\mu}$ and $\Gamma _{ij}^{\mu}$, 
for $MM\gamma$, $BB\gamma$ and $\gamma BB^{\prime}MM^{\prime}$, respectively,  are 
obtained by sum of the two contributions. 

In the minimal coupling scheme, 
the photon coupling to the meson, $V_{\trm{M}_{i}}^{\mu}$,  is given by 
\be
- i V_{\trm{M}_{i}}^{\mu}(k, \, k^{\prime})
= i Q_{\trm{M}_{i}} (k + k^{\prime})^{\mu} , 
\label{eq:Fermi-meson-photon} 
\ee
with the incoming and outgoing meson momenta $k^{\mu}$ and $k^{\prime \mu}$. 
The minimal coupling of the photon to the baryon is given by 
\begin{align}
&- i V_{\trm{B}_{i}}^{\trm{(m)}, \, \mu} (p, \, p^{\prime})
\nonumber \\
&= \left( i Q_{\trm{B}_{i}} \frac{(p + p^{\prime})^{0}}{2 M_{i}} , \, 
i Q_{\trm{B}_{i}} \frac{\bm{p} + \bm{p}^{\prime}}{2 M_{i}} + i \mu _{i}^{\trm{(N)}} 
\frac{i \bm{\sigma} \times \bm{q}}{2 M_{\trm{p}}} 
\right) ,
\label{eq:Fermi-baryon-photon}
\end{align}
with the normal magnetic moments $\mu _{i}^{\trm{(N)}}$ and 
the incoming and outgoing baryon momenta  $p^{\mu}$ 
and $p^{\prime \mu}$. 
Here we have performed non-relativistic reduction. 
These two couplings (\ref{eq:Fermi-meson-photon}) and 
(\ref{eq:Fermi-baryon-photon}) are appropriate 
with the propagators in the loop function (\ref{eq:loop}). 
The spatial components without the Pauli matrices $\bm{\sigma}$ 
in Eqs.~(\ref{eq:Fermi-meson-photon}) and 
(\ref{eq:Fermi-baryon-photon}) 
give no contribute to the $\Lambda(1405)$ form factors in the 
Breit frame. 
%
%
For the $\gamma BB^{\prime} MM^{\prime}$ coupling 
appearing in $T_{\gamma 3 ij}^{\mu =0}$, we use the following
vertex which is obtained  so that  Ward-Takahashi identity 
is satisfied with Eqs.~(\ref{eq:Vij}), (\ref{eq:Fermi-meson-photon}) and (\ref{eq:Fermi-baryon-photon}) in tree-level: 
\begin{align}
&- i \Gamma _{ij}^{(\trm{m}), \, \mu}(P, \, P^{\prime}) 
\nonumber \\ &= i \frac{C_{ij}}{4 f^{2}} 
\frac{P^{\mu} + P^{\prime \mu}}
{\sqrt{\mathstrut s} + \sqrt{\mathstrut s^{\prime}}} 
(Q_{\trm{M}_{i}} + Q_{\trm{B}_{i}} + Q_{\trm{M}_{j}} + Q_{\trm{B}_{j}}) , 
\label{eq:FermiWTphoton}
\end{align}
with incoming and outgoing meson-baryon momenta, $P^{\mu}$ and 
$P^{\prime \mu}$, respectively. Actually for the neutral excited baryon, 
this term does not contribute due to 
$Q_{\trm{H}}=Q_{\trm{B}}+Q_{\trm{M}}=0$. 

For the $BB\gamma$ and $\gamma BB^{\prime} MM^{\prime}$ couplings
we have also the anomalous coupling terms, which are 
gauge-invariant by themselves.
For these couplings, we use the interaction Lagrangian 
appearing in the chiral perturbation theory~\cite{Meissner:1997hn}:
\begin{align}
\mathcal{L}_{\trm{int}} = & - \frac{i}{4 M_{p}} b_{6}^{\trm{F}} 
\trm{Tr} \left( \overline{B} [ S^{\mu}, S^{\nu}][F_{\mu \nu}^{+}, B]
   \right) \nonumber \\ &
   -\frac{i}{4 M_{p}} b_{6}^{\trm{D}} \trm{Tr}
\left( \overline{B} [ S^{\mu}, S^{\nu}] \{ F_{\mu \nu}^{+}, B \}
   \right) , 
\label{eq:Pauliint}
\end{align}
with
\begin{equation}
   F^{+}_{\mu \nu} = - e \left( u^{\dagger} Q F_{\mu \nu} u 
+ u Q F_{\mu \nu} u^{\dagger} \right), 
\end{equation}
the electromagnetic field tensor $F_{\mu \nu}$, the charge 
matrix $Q$, the spin matrix $S_{\mu}$ and the chiral field 
$u^{2}= U = \exp(i \sqrt{\Phi} /f)$ where $\Phi$ is the SU(3) 
matrix of the Nambu-Goldstone boson field.
This interaction Lagrangian gives us spatial components of 
both the $BB\gamma$ and the 
$\gamma BB^{\prime} MM^{\prime}$ vertices:
\begin{align}
& - i V_{\trm{B}_{i}}^{\trm{(A)}, \, a} = 
i \mu^{\trm{(A)}}_{i} \left( 
\frac{i \bm{\sigma} \times \bm{q}}{2 M_{\trm{p}}} 
\right) ^{a} , \\
& - i \Gamma _{ij}^{\trm{(A)}, \, a} = A_{ij} \left(
\frac{i \bm{\sigma} \times \bm{q}}{2 M_{\trm{p}}} 
\right) ^{a} , 
\label{eq:GBa} 
\end{align}
where we have made non-relativistic reduction. 
$\mu _{i}^{\trm{(A)}}$ are the anomalous 
magnetic moments of the baryons, and the matrix $A_{ij}$ is given as, 
\begin{equation}
A_{ij} =  i \frac{b_{6}^{\trm{D}} X_{ij} + 
 b_{6}^{\trm{F}} Y_{ij}}{2 f^{2}} , 
\end{equation}
with the
coefficients 
$X_{ij}$, $Y_{ij}$ fixed only by the flavor SU(3) symmetry and 
their explicit values are found in 
Ref.~\cite{Jido:2002yz}. 
The anomalous magnetic moments are given by the interaction
Lagrangian (\ref{eq:Pauliint}) as $\mu _{i}^{\trm{(A)}} = 
b_{6}^{\trm{D}} d_{i} + b_{6}^{\trm{F}} f_{i}$
with the SU(3) coefficients, $d_{i}$ and $f_{i}$. 
The values of the coefficients $b_{6}^{\trm{D}}$ and 
$b_{6}^{\trm{F}}$ are fixed by $b_{6}^{\trm{D}}=2.40$ and 
$b_{6}^{\trm{F}}=0.82$ so as to reproduce the observed 
$\mu _{i}^{\trm{(A)}}$ of the baryons. 
In the calculation, these values are used for the 
$\gamma BB^{\prime} MM^{\prime}$ vertices (\ref{eq:GBa}), 
while for the baryon magnetic moments we use the experimental values.
For the unobserved $\Sigma ^{0}$ magnetic moment, 
we use the SU(3) flavor relation $\mu_{\Sigma ^{0}} = 
(\mu_{\Sigma ^{+}} + \mu_{\Sigma ^{-}})/2$, which
is consistent with the quark model. 
We neglect the transition magnetic moment 
$\mu_{\Lambda \Sigma ^{0}}$, 
because this interaction changes the isospin of the 
excited baryons $0$ to $1$. 

After obtaining the elementary couplings, we calculate
the amplitudes, $T_{\gamma 1 ij}^{\mu}$, 
$T_{\gamma 2 ij}^{\mu}$ 
and $T_{\gamma 3 ij}^{\mu}$, accordingly to the Feynman diagrams
given in Fig.~\ref{fig:Tgamma}: 
\begin{align}
& T_{\gamma 1 ij}^{\mu} = \sum _{l} 
T_{il}(\sqrt{\mathstrut s^{\prime}}) D_{\trm{M}_{l}}^{\mu} 
T_{lj}(\sqrt{\mathstrut s}) , 
\label{eq:TDMT} \\
& T_{\gamma 2 ij}^{\mu} = \sum _{l} 
T_{il}(\sqrt{\mathstrut s^{\prime}}) D_{\trm{B}_{l}}^{\mu} 
T_{lj}(\sqrt{\mathstrut s}) , 
\label{eq:TDBT} \\
& T_{\gamma 3 ij}^{\mu} = \sum _{k,l}
T_{ik}(\sqrt{\mathstrut s^{\prime}}) G_{k}(\sqrt{\mathstrut s^{\prime}}) 
\Gamma _{kl}^{\mu} G_{l} (\sqrt{\mathstrut s}) T_{lj}(\sqrt{\mathstrut s}) ,
\label{eq:TGGGT}
\end{align}
where the loop integrals are given with 
the photon couplings to the meson and baryon by
\begin{align}
& D_{\trm{M}_{l}}^{\mu} \equiv i
\int \frac{d^4 q_{1}}{(2 \pi )^4} 
\frac{2 M_{l}}{(P - q_{1})^2 - M_{l}^2}
\frac{1}{(q_{1} + q)^{2} - m_{l}^{2}} 
 \nonumber \\
& \phantom{0} \times 
\big[ V_{\trm{M}_{l}}^{\mu} 
(q_{1}, \, q_{1} + q) \big]
\frac{1}{q_{1}^2 - m_{l}^2} , 
\label{eq:DMloop}
\end{align}
\begin{align}
& D_{\trm{B}_{l}}^{\mu} \equiv i 
\int \frac{d^4 q_{1}}{(2 \pi )^4} 
\frac{1}{q_{1}^2 - m_{l}^2} \frac{2 M_{l}}
 {(P + q - q_{1})^2 - M_{l}^2} \nonumber \\ & 
\phantom{0} 
\times \big[ V_{\trm{B}_{l}}^{\mu} 
(P - q_{1}, \, P - q_{1} + q) \big]
\frac{2 M_{l}}{(P - q_{1})^2 - M_{l}^2} .
\label{eq:DBloop}
\end{align}
The loop function $G$ in 
Eq.~(\ref{eq:TGGGT}) is regularized in the same way as 
Eq.~(\ref{eq:loop}) with the subtraction 
constants~(\ref{eq:subtconst}).  This treatment, in fact, is necessary 
for the gauge-invariant calculation. 
The integrals in Eqs.~(\ref{eq:DMloop}) and (\ref{eq:DBloop}) 
are convergent and require no regularizations, because $V_{\trm{M}_{l}}^{\mu}$ 
and $V_{\trm{B}_{l}}^{\mu}$ are not more than second power of the 
loop momentum. 

In the present calculation, we did not introduce the form factors 
for the ground state mesons and baryons and treat them as point
particles, because we are interested in the sizes of the 
excited baryon generated by meson-baryon dynamics
and in estimation of pure dynamical effects. For the qualitative 
argument or comparison with experiments (if possible), 
the inclusion of the form factors will be important. 
In this formulation, inclusion of the form factors for the 
mesons and baryons is straightforward; 
we simply multiply the meson and baryon form factors 
to each vertex. Here, the gauge invariance requires to use a 
common form factor $F(Q^{2})$ to every vertex. The common form 
factor can be factorized out from 
the loop integrals since it depends only on the 
photon momentum, so the correction from the inclusion of the form
factor is multiplicative: $\tilde{G}(Q^{2}) = G(Q^{2}) F(Q^{2})$, 
where $\tilde{G}(Q^{2})$ and $G(Q^{2})$ are the form factors 
of excited baryons with and without the 
inclusion of the meson and baryon form factors, respectively. 
It is interesting to note that the electric mean squared 
radii for neutral excited baryons do {\it not} depend on 
the inclusion of the meson and baryon form factor, since 
\be
\EMSR \sim \tilde{G}^{\prime}_{\trm{E}}(0) = 
G^{\prime}_{\trm{E}}(0) F(0) + G_{\trm{E}}(0) F^{\prime}(0) 
\ee
and  $\GE (0)=0$ for neutral excited baryons 
and $F(0)=1$ by definition. Thus, the present results 
for $\EMSR$ of the $\Lambda (1405)$ remain unchanged even with inclusion 
of the meson and baryon form factor.

\section{Numerical results}
\label{sect:numer}

\begin{figure}[!b]
 \bc
 \begin{tabular}{c}
   \includegraphics[scale=0.3,angle=-90]{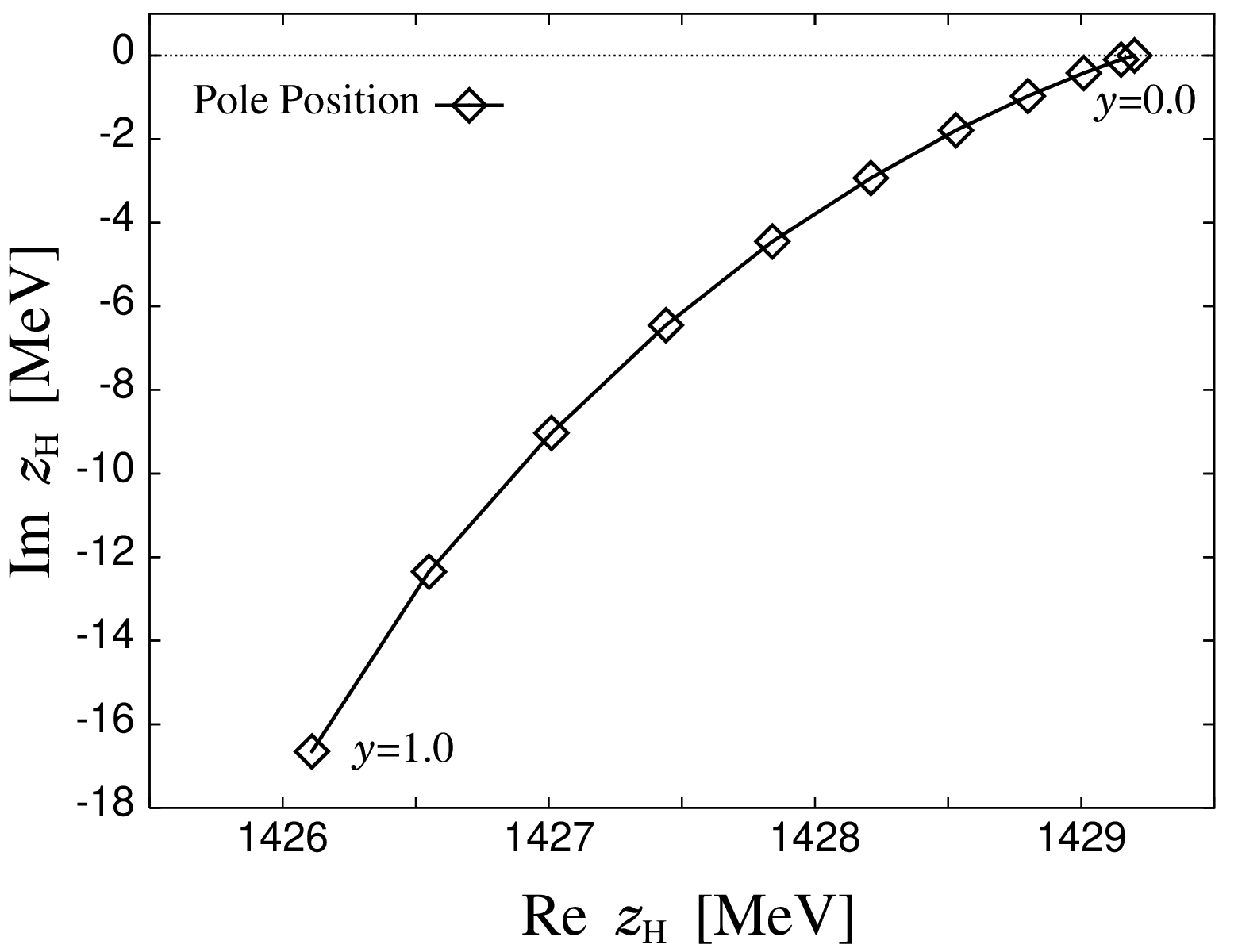} \\
   \includegraphics[scale=0.3,angle=-90]{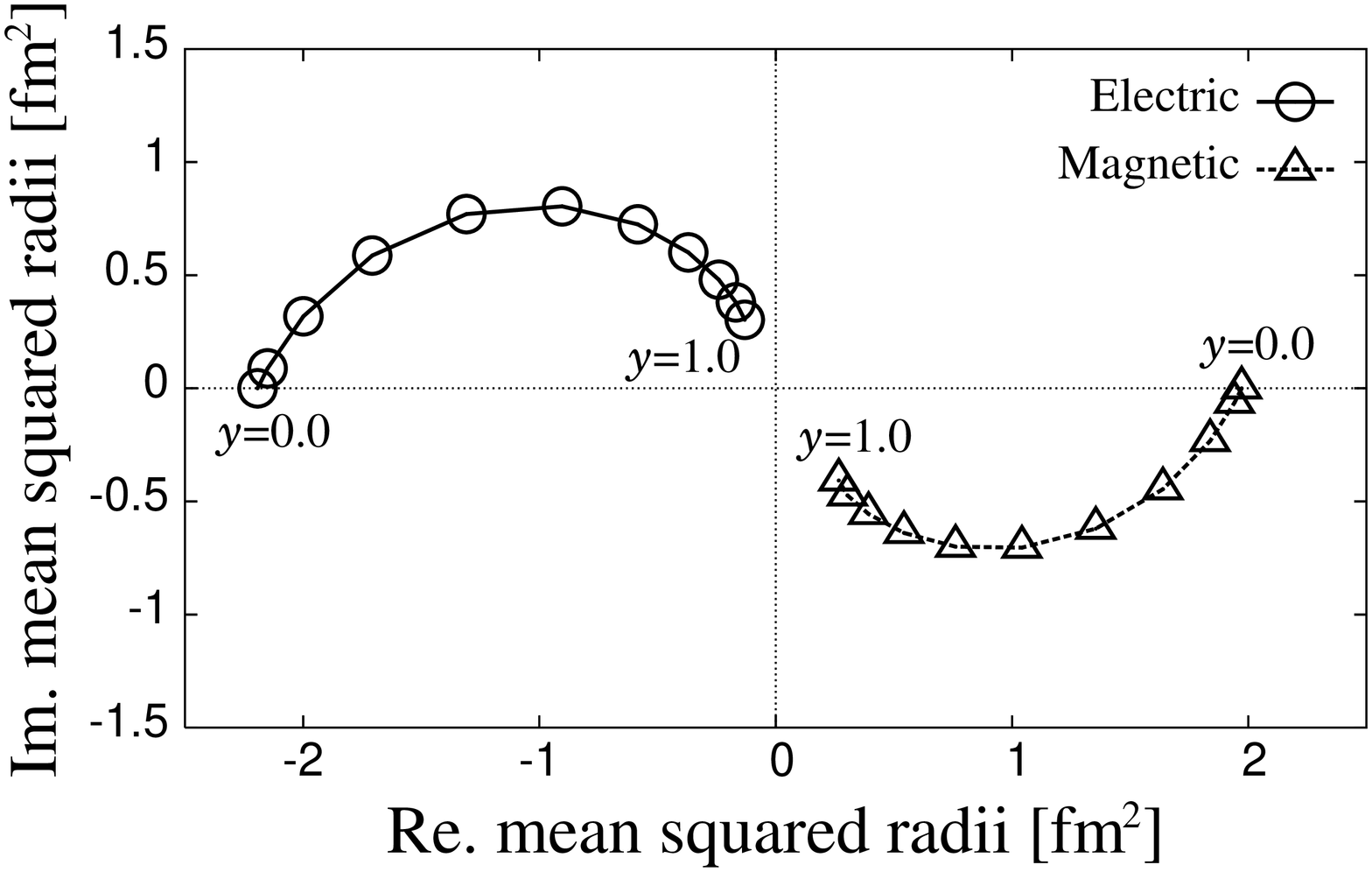}
 \end{tabular}
  \caption{Upper: Trajectory of the pole position $z_{2}$ 
of the BS amplitude by changing the off-diagonal couplings 
of $\KbarN$ to other channels. 
Lower: Trajectory of the electromagnetic mean 
squared radii of the $z_{2}$. Both 
are in step of $0.1$ for the parameter $y$. }
  \label{fig:KN}
 \ec
\end{figure}

\begin{table*}[!ht]
\caption{Electromagnetic mean squared radii, $\EMSR$ and $\MMSR$, 
 of $\Lambda (1405)$.}
\label{tab:results} 
\begin{tabular*}{\textwidth}{@{\extracolsep{\fill}}ccccc}
 \toprule
State & Pole position [MeV] & Strongly couple to &
 $\EMSR$ [$\trm{fm}^{2}$] &
 $\MMSR$ [$\trm{fm}^{2}$] \\
   \midrule
   $z_{1}$
   & 1390.43 $-$ 66.21$i$
   & $\pi \Sigma$ 
   & $\phantom{-} (1.8 - 0.2 i ) \times 10^{-2}$ 
   & $(-1.3 + 2.1 i) \times 10^{-2}$ \\
   $z_{2}$
   & 1426.11 $-$ 16.65$i$
   & $\KbarN$ 
   & $-0.131 + 0.303 i$ 
   & $\phantom{-} 0.267 - 0.407 i \phantom{-}$ \\
\midrule
   No decay to $\pi \Sigma$
   & 1422.34  $-$ \phantom{0}0.02$i$ 
   & $\Kmp$, $\Kzn$ 
   & $-0.519 - 0.008 i$ 
   & $\phantom{-} 0.683 - 0.023 i \phantom{-}$ \\
   $\KbarN$ bound state 
   & 1429.20\phantom{ $-$ 16.65$i$} 
   & $\Kmp$, $\Kzn$ 
   & $-2.193$ \phantom{$- 0.000 i $} 
   & $\phantom{+} 1.972$ \phantom{$+ 0.000 i \phantom{-}$} \\
   \bottomrule
\end{tabular*}
\end{table*}

We show our result for the electromagnetic mean squared 
radii of the excited baryons in Table~\ref{tab:results}. 
As mentioned before, we have two $\Lambda (1405)$ states, $z_{1}$
and $z_{2}$.
The lower state strongly couples to the 
$\pi \Sigma$ channel, while the higher state dominantly 
couples to the $\KbarN$ channel, as one can see from the 
analysis of $g_{i}$ in Ref.~\cite{Jido:2003cb}.
Since the $\pi^{+}\Sigma^{-}$ and $\pi^{-}\Sigma^{+}$ 
contribute to the isospin $0$ state almost equally, 
the electric mean squared radius of the lower $\Lambda (1405)$ state 
$z_{1}$ is suppressed. 

The electric mean squared radius of the higher $\Lambda (1405)$ 
state $z_2$ is more interesting. The higher state has 
a three times larger absolute value of 
$\EMSR$, $0.32 \, \trm{fm}^{2}$, than that of 
neutron $\sim - 0.12 \, \trm{fm}^{2}$.
This observation implies that the electric form factor of the $\Lambda(1405)$
is softer, namely has larger energy dependence,  
than that of the neutron, since the electric form factors for the
neutral particles are given by $G_{\trm{E}}(Q^{2}) = - Q^{2} 
\EMSR/6 + \cdots$ in the expansion of $Q^{2}$ due to the 
neutral charge $G_{\trm{E}}(0)=0$. 
The softer form factor means that the $\Lambda (1405)$ is
more spatially extended. This is because kaon inside the
$\Lambda (1405)$ has less ``virtuality'' than pions surrounding
the nucleon~\cite{Yamazaki:2007fk}. The negative charge radius
of the neutron is interpreted as distribution of $\pi ^{-}$ cloud. 
For the neutron case, the pion cloud consists of completely virtual pions, 
since the system needs at least 140 MeV to create a pion. 
In contrast, for the $z_{2}$ pole of the $\Lambda (1405)$,
only several MeV is required to make $\bar{K}N$ state
which is the dominant component of this resonance
according to the analysis of the coupling strengths~\cite{Jido:2003cb}.
Therefore, the $K^{-}$ inside the $\Lambda (1405)$ can be largely 
distributed. 

These results are free from the non-resonant background 
since we calculate the mean squared radii
on the top of the $\Lambda(1405)$ resonance pole 
in the complex energy plane. Thus, these values have definite 
theoretical meanings and are used for comparison with other 
models. On the other hand, experimental observables may be 
the ratio of the amplitudes 
$T_{\gamma ij}^{\mu }/T_{ij}$ in the real energies,
which includes the contributions from the non-resonant 
meson-baryon scattering states~\cite{Jido:2002yz}. 

In order to extract the information of the sizes from the 
form factors, we perform the following analysis. If the decay 
width of the resonance is small, the imaginary parts of the 
mean squared radii are small. Thus, 
it is possible to interpret the mean squared radii as the sizes 
for the resonant states, since the radii 
are close to real numbers. 
For this purpose, we analyze the mean squared radii
of a $\Lambda(1405)$ calculated without the $\pi \Sigma$
channel, which is the main decay mode of the actual $\Lambda(1405)$. 
Neglecting the couplings to the $\pi \Sigma$ channel in $C_{ij}$ and 
leaving other parameters as in the calculation of the actual 
$\Lambda(1405)$,
we find the pole position $1422.34 - 0.02 i \, \trm{MeV}$, the mean 
squared radii $\EMSR = -0.519 - 0.008 i \, \trm{fm}^{2}$ 
and $\MMSR = 0.683 - 0.023 i \, \trm{fm}^{2}$. 
The small imaginary parts of these values come from 
the coupling to the open $\pi \Lambda$ channel through 
the tiny isospin breaking
in the masses of the constituent baryons and mesons. 
The absolute values of $\EMSR$ and $\MMSR$ are very 
similar to those obtained
by the actual $\Lambda(1405)$ of the $z_{2}$ pole. 

We also perform analyses for another interesting system of  
a $\bar KN$ bound state, which is generated only by the attractive 
interaction of the $\KbarN$ channel~\cite{Hyodo:2007jq}. 
This bound state is considered to be an origin of the higher 
state of the actual $\Lambda (1405)$, $z_{2}$. 
To investigate this state, we introduce a parameter $y$ ($0 \leq y \leq 1$)
in front of the off-diagonal couplings of $\KbarN$ to the other channels 
in $C_{ij}$. The case of $y=1$ corresponds to the full coupled-channel calculation, 
whereas $y=0$ corresponds to the calculation with only the $\KbarN$ channels.

\begin{figure}[b]
 \includegraphics[scale=0.3,angle=-90]{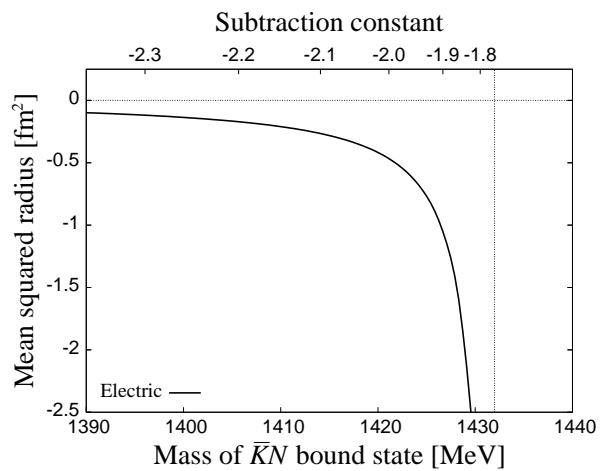}
 \caption{Electric mean squared radius of the $\KbarN$ bound 
state as a function of the mass. The upper horizontal axis denotes 
the corresponding values of the subtraction constants $a_{\KbarN}$. 
The vertical dotted line represents position of the $\Kmp$ threshold.}
\label{fig:RvsB}
\end{figure}

The trajectory of the pole position from $y=1$ to $y=0$ is shown in 
Fig.~\ref{fig:KN}. We also show the corresponding electromagnetic mean squared 
radii in Fig.~\ref{fig:KN}.
We thus estimate the size of a meson-baryon resonance by
the $\KbarN$ bound state  which appears 
at $1429 \, \trm{MeV}$ in our model (see Tab.~\ref{tab:results}). 
The bound state consists of $\Kmp$ and $\Kzn$ 
components. The electric mean squared radius reflects the 
charge distribution of the $\Kmp$ component, since the 
$\Kzn$, in which both hadrons are charge neutral,  
does not contribute to the electric interactions. 
The negative sign for the radius implies that the 
$K^{-}$ is surrounding around the proton. 
In addition, with the fact that the electromagnetic size
of the proton is roughly $0.9 \, \trm{fm}$, 
our result of the electric root mean squared radius 
$\sqrt{| \EMSR |} \simeq 1.48 \, \trm{fm}$ implies 
that $\Lambda (1405)$ has structure of widely spread $K^{-}$ 
clouds around the core of proton with larger size than that of typical 
ground state baryons. The magnetic mean squared radius, $\MMSR$, 
represents  distribution of magnetic moments of the core nucleons in 
the $\bar KN$ bound system,  because anti-kaons cannot contribute to 
magnetic interactions. Due to dynamics of the nucleon and 
anti-kaon, the nucleon is distributed  inside of the $\Lambda(1405)$,
of which size is seen as the mean squared radii. This implies that
the magnetic interaction also suggests a larger size for the $\Lambda(1405)$
than the ordinary baryons. 

The results of the mean squared radii of the $\Lambda(1405)$ 
without the $\pi \Sigma$ decay channel are much smaller than those
of the $\KbarN$ bound state.
This is because the former case provides deeper bound state than the latter
case. The resonance energies of these cases are very similar as seen by  
just 7 MeV difference out of about 1 GeV energy scale, but the binding 
energies have two times difference. 
To investigate the relation between the radii and the binding energy,
we consider the $\KbarN $ bound state and tune the 
binding energy by changing 
the subtraction constant $a_{\KbarN}$, which is the only 
parameter of this system. In Fig.~\ref{fig:RvsB}, 
we show the electric mean squared 
radius as a function of the mass of the $\KbarN$ bound state. 
This figure implies that the deeper bound states have the 
smaller radii. Since the smaller size of the bound state is 
expected in the deeper bound state, the obtained radius can be 
interpreted as the size of the $\Lambda (1405)$. 
Therefore, knowing precise resonance point of the $\Lambda(1405)$ is
very important to understand its properties.

Let us examine our result of electric mean squared radius in 
comparison with theoretical estimates of the size of the $\Lambda (1405)$. 
In the study of the kaonic nuclei, single-channel $\KbarN$ 
potentials were constructed from the phenomenological 
interaction~\cite{Yamazaki:2007cs} and from the chiral 
coupled-channel approach~\cite{Hyodo:2007jq}, 
both of which reproduce experimental data of $\KbarN$ 
phenomenology. The relative mean distance of $\bar{K}$ and 
$N$ in $\Lambda (1405)$ was estimated to be about 
$1.36 \, \trm{fm}$ by phenomenological 
potential~\cite{Yamazaki:2007cs} and about $1.8 \, \trm{fm}$ 
by chiral potential~\cite{Dote:2008in} where the difference of 
the results is attributed to the difference of binding energies 
and strengths of the potentials. Our result, 
$\sqrt{| \EMSR |} \simeq 1.48 \, \trm{fm}$, is 
quantitatively similar to the estimated size of the $\Lambda (1405)$, 
although one should note that we have evaluated the charge 
radius which cannot be directly compared with the mean 
distance of $\bar{K}$ and $N$.

\section{Conclusions}
\label{sect:conclusions}

We have calculated the electromagnetic mean squared radii 
of $\Lambda (1405)$ based on the meson-baryon picture in the 
chiral unitary model. The evaluation of the electromagnetic 
mean squared radii have been performed in two ways: In the 
first approach, with full coupled channels for the $\Lambda (1405)$, 
the mean squared radii in complex numbers are obtained at 
the poles for the physical resonant states and the absolute 
values suggest that the form factors
for the $\Lambda (1405)$ is softer than
those for the neutron. In the second, we 
describe $\Lambda (1405)$ as a bound state of $\KbarN$ by 
neglecting all the off-diagonal couplings of $\KbarN$ to the 
other channels, in order to estimate the size of the resonant 
state. As a consequence of the small binding energy in chiral 
unitary model, our result implies that $K^{-}$ in $\Lambda (1405)$ is 
widely spread around $p$ and that the size of the 
$\Lambda (1405)$ is larger than that of typical ground state baryons. 

\vspace{0.5cm}

\noindent {\bf Acknowledgments} 

\vspace{0.5cm}


We acknowledge Y.~Kanada-En'yo, T.~Myo and A.~Hosaka for 
usefull discussions. T.S. thanks H.~Suganuma for 
continuous encouragement. T.H. thanks the Japanese Society for the 
Promotion of Science for financial support. 
This work is partly supported by 
the Grant for Scientific Research (No. 19853500, No. 18042001 
and No. 20028004)
and 
the Grant-in-Aid for the 21st Century COE ``Center for Diversity and
Universality in Physics'' from the Ministry of Education, Culture,
Sports, Science and Technology of Japan.
This work was
done under Yukawa International Program for Quark-Hadron Sciences.

\vspace{0.5cm}

\noindent {\bf Appendix}

\vspace{0.5cm}


In this appendix, we show that the present formulation for
the amplitude with electromagnetic coupling developed in 
Sec.~\ref{sect:photon} and \ref{sect:EvalFF} satisfies 
Ward-Takahashi identity at the resonance point of the 
excited baryon with charge $Q_{\trm{H}}$.

The photon coupling amplitude in the present formulation is given by three contributions:
\begin{equation}
 T_{\gamma}^{\mu} = 
 T_{\gamma 1}^{\mu} + T_{\gamma 2}^{\mu} + T_{\gamma 3}^{\mu} .
\end{equation}
The definition of each term is, as given in Eqs.~(\ref{eq:TDMT}), 
(\ref{eq:TDBT}) and (\ref{eq:TGGGT}),
\begin{align}
& T_{\gamma 1 ij}^{\mu} = \sum _{l} 
T_{il}(\sqrt{\mathstrut s^{\prime}}) D_{\trm{M}_{l}}^{\mu} 
T_{lj}(\sqrt{\mathstrut s}) , 
\nonumber \\
& T_{\gamma 2 ij}^{\mu} = \sum _{l} 
T_{il}(\sqrt{\mathstrut s^{\prime}}) D_{\trm{B}_{l}}^{\mu} 
T_{lj}(\mathstrut \sqrt{s}) , 
 \\
& T_{\gamma 3 ij}^{\mu} = \sum _{k,l}
T_{ik}(\sqrt{\mathstrut s^{\prime}}) G_{k}(\sqrt{\mathstrut s^{\prime}}) 
\Gamma _{kl}^{\mu} 
G_{l}(\sqrt{\mathstrut s}) T_{lj}(\sqrt{\mathstrut s}) , \nonumber 
\end{align}
where $s \equiv P^{\mu}P_{\mu}$, $s^{\prime} \equiv P^{\prime \mu}P_{\mu}^{\prime}$ 
and $P^{\prime \mu}=P^{\mu}+q^{\mu}$ with the photon momentum $q_{\mu}$.
Let us also recall the photon couplings appearing these three contributions:
\begin{align}
&- i V_{\trm{M}_{i}}^{\mu}(k, \, k^{\prime})
= i Q_{\trm{M}_{i}} (k + k^{\prime})^{\mu} , 
\nonumber \\
&- i V_{\trm{B}_{i}}^{\mu}(p, \, p^{\prime})
= i Q_{\trm{B}_{i}} (p + p^{\prime})^{\mu} , \\
&- i \Gamma _{ij}^{\mu} (P, \, P^{\prime}) = i Q_{\trm{H}} 
\frac{C_{ij}}{2 f^{2}} \frac{P^{\mu} + P^{\prime \mu}}
{\sqrt{\mathstrut s} + \sqrt{\mathstrut s^{\prime}}} ,
\nonumber 
\end{align}
with incoming and outgoing momenta of mesons $k^{\mu}$, $k^{\prime \mu}$ 
and baryons $p^{\mu}$, $p^{\prime \mu}$, charge of mesons $Q_{\trm{M}_{i}}$ 
and baryons $Q_{\trm{B}_{i}}$, and $P^{\mu}=k^{\mu}+p^{\mu}$, 
$P^{\prime \mu}=k^{\prime \mu}+p^{\prime \mu}$, $Q_{\trm{H}}=Q_{\trm{M}}+Q_{\trm{B}}$. 
Here we have omitted the terms with the Pauli matrices $\bm{\sigma}$, 
which vanish when $q^{\mu}$ is multiplied. 

With the aid 
of the following identity,
\begin{align}
& q_{\mu} \frac{1}{(q_{1} + q)^{2} - m^{2}} (2 q_{1} + q)^{\mu} 
\frac{1}{q_{1}^{2} - m^{2}} \nonumber \\
& = \frac{1}{q_{1}^{2} - m^{2}}
- \frac{1}{(q_{1} + q)^{2} - m^{2}} , 
\end{align}
we can obtain the loop integrals in Eqs~(\ref{eq:DMloop}) and 
(\ref{eq:DBloop}) in terms of the loop function $G$:
\be
\begin{split}
& q_{\mu} D_{\trm{M}_{l}}^{\mu} = Q_{\trm{M}_{l}} 
\left[ G_{l}(\sqrt{\mathstrut s^{\prime}}) - 
G_{l}(\sqrt{\mathstrut s}) \right] ,
\\
& q_{\mu} D_{\trm{B}_{l}}^{\mu} = Q_{\trm{B}_{l}} 
\left[ G_{l}(\sqrt{\mathstrut s^{\prime}}) - 
G_{l}(\sqrt{\mathstrut s}) \right] .
\end{split}
\ee 
Therefore, multiplying $q_{\mu}$ to 
$T_{\gamma 1 ij}^{\mu}+T_{\gamma 2 ij}^{\mu}$, we obtain
\be
q_{\mu} \left( T_{\gamma 1 ij}^{\mu} + T_{\gamma 2 ij}^{\mu} \right)
= Q_{\trm{H}} \sum _{l} T_{il}^{\prime} 
\big[ G_{l}^{\prime} - G_{l} \big] T_{lj} .
\ee
Here $G^{\prime}$ and $T^{\prime}$ mean $G(\sqrt{s^{\prime}})$ and $T(\sqrt{s^{\prime}})$, respectively. 

The coupling $\Gamma _{ij}^{\mu}$ satisfies  
\begin{align}
q_{\mu} \Gamma _{ij}^{\mu}(P, \, P^{\prime}) & = - Q_{\trm{H}} 
\frac{C_{ij}}{2 f^{2}} 
\left( \sqrt{\mathstrut s^{\prime}} - \sqrt{\mathstrut s} \right)
\nonumber \\
& = Q_{\trm{H}} \big[ V_{ij}^{\prime} - V_{ij} \big] ,
\end{align}
with the Weinberg-Tomozawa interaction $V_{ij}^{\prime} \equiv V_{ij}(\sqrt{s^{\prime}})$ and 
$V_{ij} \equiv V_{ij}(\sqrt{s})$. Thus, 
\be
q_{\mu} T_{\gamma 3 ij}^{\mu} 
= Q_{\trm{H}} \sum _{k, l} T_{ik}^{\prime} 
G_{k}^{\prime} \big[ V_{kl}^{\prime} - V_{kl} \big] G_{l} T_{lj} .
\ee

Collecting all the terms and using BS 
equation~(\ref{eq:BSEq}), we obtain 
\begin{align}
q_{\mu} T_{\gamma}^{\mu} &= 
Q_{\trm{H}} \big[ T^{\prime} G^{\prime} \left(T - VGT \right) 
- \left( T^{\prime} - T^{\prime} G^{\prime} V^{\prime} \right) G T \big] 
\nonumber \\
&= Q_{\trm{H}} \big[ T^{\prime} G^{\prime} V - V^{\prime} G T \big] , 
\end{align}
in matrix form. This equation has only single pole terms in the 
right-hand side. 
For calculation of the form factor of the baryon resonances,  
we extract residues of double pole of the amplitude in the Breit frame
and the right-hand side of this equation does not contribute to 
the residues. Therefore, 
the present formulation satisfies Ward-Takahashi 
identity at the resonance position and the form factors evaluated by 
$T_{\gamma}^{\mu}$ are gauge-invariant.
It is also seen in the above argument that, off the pole position, 
it is necessary to include the other terms shown in 
Ref.~\cite{Borasoy:2005zg} for gauge invariance.


\begin{thebibliography}{99}

\bibitem{Dalitz:1960du}
  R.~H.~Dalitz and S.~F.~Tuan,
  Ann. Phys.\  {\bf 10} (1960) 307 .

\bibitem{Dalitz:1967fp}
  R.~H.~Dalitz, T.~C.~Wong and G.~Rajasekaran,
  Phys.\ Rev.\  {\bf 153} (1967) 1617.

\bibitem{Wyld:1967}
  J.~H.~W.~Wyld, 
  Phys.\ Rev.\  {\bf 155} (1967) 1649.

\bibitem{Kaiser:1995eg}
  N.~Kaiser, P.~B.~Siegel and W.~Weise,
  Nucl.\ Phys.\  A {\bf 594} (1995) 325.

\bibitem{Oset:1997it}
  E.~Oset and A.~Ramos,
  Nucl.\ Phys.\  A {\bf 635} (1998) 99.


\bibitem{Oller:2000fj}
  J.~A.~Oller and U.~G.~Meissner,
  Phys.\ Lett.\  B {\bf 500} (2001) 263.

\bibitem{Oset:2001cn}
  E.~Oset, A.~Ramos and C.~Bennhold,
  Phys.\ Lett.\  B {\bf 527} (2002) 99.

\bibitem{Lutz:2001yb}
  M.~F.~M.~Lutz and E.~E.~Kolomeitsev,
  Nucl.\ Phys.\  A {\bf 700} (2002) 193.

\bibitem{Jido:2003cb}
  D.~Jido, J.~A.~Oller, E.~Oset, A.~Ramos and U.~G.~Meissner,
  Nucl.\ Phys.\  A {\bf 725} (2003) 181.

\bibitem{Hyodo:2008xr}
  T.~Hyodo, D.~Jido and A.~Hosaka,
  Phys.\ Rev.\ C {\bf 78} (2008) 025203.

\bibitem{Jido:2002zk}
  D.~Jido, E.~Oset and A.~Ramos,
  Phys.\ Rev.\  C {\bf 66} (2002) 055203.
  
\bibitem{Fink:1989uk}
In a different context, the two poles for the $\Lambda (1405)$ was discussed in 
  J.~P.~Fink~Jr., G.~He, R.~H.~Landau and J.~W.~Schnick,
  Phys.\ Rev.\  C {\bf 41} (1990) 2720.

\bibitem{Kaiser:1996js}
  N.~Kaiser, T.~Waas and W.~Weise,
  Nucl.\ Phys.\  A {\bf 612} (1997) 297.

\bibitem{Nacher:1999ni}
J.~C.~Nacher, E.~Oset, H.~Toki and A.~Ramos,
Phys.\ Lett.\ B {\bf 461} (1999) 299; B {\bf 455} (1999) 55.

\bibitem{Jido:2002yz}
  D.~Jido, A.~Hosaka, J.~C.~Nacher, E.~Oset and A.~Ramos,
  Phys.\ Rev.\  C {\bf 66} (2002) 025203.

\bibitem{Geng:2007hz}
L.~S.~Geng, E.~Oset and M.~Doring,
Eur. Phys. J. A {\bf 32} (2007) 201.

\bibitem{Isgur:1978xj}
  N.~Isgur and G.~Karl,
  Phys.\ Rev.\  D {\bf 18} (1978) 4187.

\bibitem{Hyodo:2007np}
  T.~Hyodo, D.~Jido and L.~Roca,
  Phys.\ Rev.\  D {\bf 77} (2008) 056010;
  L.~Roca, T.~Hyodo and D.~Jido,
  Nucl.\ Phys. A {\bf 809} (2008) 65.
  
\bibitem{Akaishi:2002bg}
  Y.~Akaishi and T.~Yamazaki,
  Phys.\ Rev.\  C {\bf 65} (2002) 044005.

\bibitem{Yamazaki:2007cs}
  T.~Yamazaki and Y.~Akaishi,
  Phys.\ Rev.\  C {\bf 76} (2007) 045201.

\bibitem{Hyodo:2007jq}
  T.~Hyodo and W.~Weise,
  Phys.\ Rev.\  C {\bf 77} (2008) 035204.

\bibitem{Dote:2008in}
  A.~Dote, T.~Hyodo and W.~Weise,
 Nucl.\ Phys.\ A {\bf 804} (2008) 197.

\bibitem{Exp:2007}
  For instance, 
J.~K.~Ahn [LEPS Collaboration], Nucl. Phys. A {\bf 721} (2003) 715;
H.~Fujimura [LEPS TPC Collaboration], Prog. Theor. Phys. Suppl. 
 {\bf 168} (2007) 123.

\bibitem{Gasser:1984gg}
  J.~Gasser, H.~Leutwyler, 
  Nucl.\ Phys.\ B {\bf 250} (1985) 465.
  
\bibitem{Pich:1995bw}
  A.~Pich, Rep.\ Prog.\ Phys.\ {\bf 58} (1994) 563.

\bibitem{Borasoy:2005ie}
  B.~Borasoy, R.~Nissler and W.~Weise,
  Eur.\ Phys.\ J.\  A {\bf 25} (2005) 79.

\bibitem{Borasoy:2005zg}
  B.~Borasoy, P.~C.~Bruns, U.~G.~Meissner, R.~Nissler,
  Phys.\ Rev.\ C {\bf 72} (2005) 065201.

\bibitem{Meissner:1997hn}
  U.~G.~Meissner and S.~Steininger,
  Nucl.\ Phys.\  B {\bf 499} (1997) 349.

\bibitem{Yamazaki:2007fk}
  T.~Yamazaki and Y.~Akaishi,
  Proc. Jpn. Acad., Ser. B {\bf 83} (2007) 144.
  
\end{thebibliography}
\end{document}